# Measurement and Modeling of Structure-Induced Surface Scattering on Terahertz Channel

Peian Li, Yapeng Ge, Jiacheng Liu, Wenbo Liu, Jiayuan Cui, Jiabiao Zhao, Qiang Niu, Yuping Yang, Xiangzhu Meng, Yiming Zhao, Jianjun Ma

*Abstract*—**As terahertz (THz) frequencies emerge as promising candidates for next-generation wireless networks, accurate characterization of propagation mechanisms in indoor/outdoor environments becomes essential for system design and performance optimization. This article presents an experimental and theoretical investigation of structure-induced indoor surface scattering on THz channels, examining how material properties and structural configurations jointly govern channel power and angular distribution. Six representative indoor surfaces are characterized, revealing that intrinsic structural inhomogeneity - particularly the quasi-periodic earlywood-latewood arrangement in pine wood - induces measurable angular scattering whose dominant lobes and angular shifts are reproduced by a beam-propagation modeling (BPM) framework. Material-covered surface configurations are further investigated, demonstrating that thin dielectric covering layers can substantially modify reflection characteristics through thickness- and frequency- dependent thin-film interference effects. Wide-angle bistatic measurements in a conference room show that folded curtains broaden the measured angular distribution of detectable received power. These findings suggest that structure-induced surface scattering mechanisms may provide scattering-assisted propagation opportunities for indoor non-line-of-sight THz scenarios.**

This work was supported in part by the National Natural Science Foundation of China under Grant 62471033 and in part by the Special Program Project for Original Basic Interdisciplinary Innovation under the Science and Technology Innovation Plan of Beijing Institute of Technology under Grant 2025CX11010. (*Corresponding author: Jianjun Ma.*)

## I. Introduction

The future demand for ultra-high data rates has intensified the exploration of terahertz (THz) frequencies for next-generation wireless networks [1], necessitating accurate characterization and predictive modeling of propagation mechanisms in realistic deployment environments [2, 3]. At THz frequencies, the wavelength becomes comparable to the characteristic dimensions of surface features and material inhomogeneities commonly encountered in indoor settings [4, 5], rendering THz channels highly sensitive to surface textures, dielectric properties, and geometric configurations - giving rise to intricate reflection and scattering phenomena that fundamentally govern channel characteristics [6]. Indoor THz channel propagation is predominantly determined by its interactions with environmental surfaces, particularly for non-line-of-sight (NLoS) transmission paths when direct propagation is obstructed. Reconfigurable intelligent surfaces (RIS) and metasurface-based structures have been proposed to actively control channel propagation through phase and amplitude manipulation [7, 8], demonstrating beam steering and coverage enhancement capabilities [9, 10]. However, such engineered reflective structures typically entail increased system complexity, elevated fabrication costs [11], and limited robustness to environmental variations.

An alternative research direction focuses on understanding and exploiting material-dependent THz channel propagation [12]. Experimental and theoretical investigations have demonstrated that material composition and surface characteristics, particularly roughness, can significantly alter channel scattering patterns and power redistribution [13]. Measurement-based studies have quantified the reflection, transmission, and scattering coefficients of common building materials, providing essential empirical inputs for THz channel modeling [5, 14]. In parallel, scattering-aware modeling frameworks capture the frequency- and angle-dependent interactions between THz channels and material surfaces, thereby enhancing propagation prediction accuracy in complex environments [15, 16]. Accordingly, leveraging the intrinsic reflection and scattering properties of existing environmental surfaces - such as walls, ceilings, and other architectural elements - offers a low-complexity, deployment-friendly approach for constructing robust NLoS

TABLE I
MEASUREMENT AND MODELING OF COMMON BUILDING MATERIALS

| Frequency | Method | Sample | Contributions | Related research gap | Novelty of this work | Ref. |
|---|---|---|---|---|---|---|
| 97.05 GHz | Continuous wave | Basswood sheets, gypsum sheets, cement mortar sheets | Water-content and defect detection | Moisture/defect-induced variations without intrinsic structural effects | Structure-induced THz scattering | [23] |
| 113-170 GHz | | Glass door | Transmission loss, reflectivity | Limited material categories and structural diversity | Diverse materials with structure-induced scattering | [17] |
| 0.2-3 THz | Terahertz time-domain spectroscopy | Ceramic, gypsum plasterboard, polymethyl methacrylate, soda-lime float glass | Dielectric parameters, attenuation coefficient, refractive index, electrical conductivity | Dielectric properties without heterogeneity-scattering correlation | Structure-dependent scattering analysis | [18] |
| 200-500 GHz | | Glass, concrete, granite | Complex permittivity and propagation simulation | Bulk-property modeling without structural scattering | Structure-induced scattering modeling | [19] |
| 300 GHz | | Bricks, stone, cement, gypsum, wood | Scattering coefficients, refractive indices | Rough-surface model limited for complex structures | Scattering from internal quasi-periodic inhomogeneity | [5] |
| 0.2-1.2 THz | | Corroded steel plates | Absorption coefficient, refractive index, and corrosion-layer thickness | Corrosion-specific characterization with limited material scope | Material-dependent scattering across diverse surfaces | [24] |
| 100-400 GHz | Terahertz frequency-domain spectroscopy | Glass, ceiling board | Dielectric-property measurement and model fitting | Dielectric modeling without structural scattering | Structure-dependent scattering behavior | [25] |
| 219-224 GHz | Vector network analyzer | Glass, granite, polyvinyl chloride board, ceramic, wooden board | Transmission and reflection coefficients, complex permittivity | Reflection/transmission without structural-origin analysis | Structure-induced scattering mechanisms | [14] |
| 140-210 GHz | | Plexiglass, nylon | Complex permittivity characterization method | Low-loss slab characterization without heterogeneity effects | Angular scattering from structural inhomogeneity | [20] |
| 282-290 GHz | | Glass, polyvinyl chloride board, marble, plaster, raw wood | Complex permittivity, reflection and transmission coefficients | Specular characterization without off-axis scattering | Structure-induced off-axis scattering | [26] |

THz links without dedicated engineered reflectors [17].

Common indoor construction materials, including drywall, wallpaper, plywood, and PVC panels, exhibit diverse electromagnetic responses arising from variations in microstructure, dielectric properties, and surface roughness, which can significantly influence channel degradation, angular scattering distributions, and multipath characteristics [18-20]. Accurate measurement and modeling of material electromagnetic parameters and reflection characteristics across the terahertz spectrum are therefore widely regarded as essential prerequisites for realistic channel modeling and system performance evaluation. To support such efforts, various measurement techniques operating in either the time domain or the frequency domain have been employed to characterize key electromagnetic parameters relevant to THz channel propagation, with representative studies summarized in Table I. Building upon these, empirical reflection and scattering data have been integrated into ray-tracing frameworks and statistical channel models [21, 22], enabling more accurate reconstruction of multipath performance and spatial power distribution in both indoor and urban scenarios.

Collectively, these studies have established a strong foundation for characterizing bulk electromagnetic properties and material-dependent transmission, reflection, and rough-surface scattering behaviors. Nevertheless, systematic characterization and modeling of structure-induced THz angular scattering remain limited. In realistic indoor environments, internal quasi-periodic inhomogeneity, layered coverings, and folded fabrics introduce distinct structural features whose scattering mechanisms may not be fully described by conventional roughness-based models. Table I summarizes representative studies from these perspectives and highlights the remaining research gaps and the contributions of this work.

In this work, we try to investigate angular scattering that arises from the structure of common indoor surfaces rather than from their random roughness, a mechanism not addressed by the bulk-permittivity and rough-surface characterizations. We target three common indoor structure classes - internal quasi-periodic inhomogeneity, smooth/rough metallic substrates covered by thin lossy dielectrics, and room-scale folded fabric elements. For each class, angle-resolved measurements are paired with a physics-based model, and the model-measurement agreement is evaluated quantitatively. We propose a beam propagation method (BPM) calculation parameterized by the growth-layer period, tilt angle, and duty cycle for the first class, a thin-film interference model using the measured complex permittivity for the second, and a physical-optics (PO) model using the measured fold geometry for the third. Unlike our earlier studies of rough [13] and wavy metallic [9] surfaces, where scattering is governed by the surface height profile, this work aims to relate the angular scattering features to identifiable structural parameters of the surfaces. The remainder of this article is organized as follows: Section II presents material characterization and investigates structure-induced scattering arising from intrinsic material inhomogeneity, focusing on the effects of pine growth-layer period, tilt angle, and duty cycle. Section III examines scattering behaviors of material-covered surfaces, highlighting thin-film interference and angular power redistribution induced by covering layers. Section IV reports wide-angle indoor measurements that reveal environment-induced scattering effects and angular spreading originating from structured indoor elements in realistic propagation scenarios. Finally, Section V summarizes the key findings.

## II. CHARACTERIZATION OF SINGLE-MATERIAL SURFACES

### A. Samples Preparation and Characterization

Realistic indoor environments contain a wide variety of materials, including decorative coatings and soft furnishings [27], whose electromagnetic and structural characteristics can differ markedly from conventional rigid surfaces. To investigate material-dependent THz channel propagation in more representative scenarios, six commonly encountered materials - wall plaster, leather, PVC leather, wallpaper, pine wood, and curtain - were selected. As illustrated in Fig. 1(a), these materials exhibit varying surface textures and can be considered electromagnetically smooth at low-THz frequencies [28], since their surface roughness remains substantially smaller than the operating wavelength according to the Rayleigh roughness criterion [4].

The tested samples were collected from realistic indoor environments and common furnishings, with thickness variations evaluated through measurements at multiple locations within the illuminated region. Among these materials, the pine wood sample is of interest due to its intrinsic structural inhomogeneity arising from the alternating arrangement of earlywood (low-density) and latewood (high-density) regions, as illustrated in Fig. 1(b). This periodic density variation, characterized by the effective period $G$ along the orientation direction $s$, is a feature of softwood species that influences electromagnetic interactions at THz frequencies [29, 30]. Despite these structural and thickness variations, accurate electromagnetic parameters were extracted using the complex transmission response of a plane-parallel dielectric slab [31], representing the effective electromagnetic response of the characterized samples under the employed THz-TDS conditions.

Fig. 1(c) and Fig. 1(d) present the extracted real ($\varepsilon'$) and imaginary ($\varepsilon''$) parts of the complex relative permittivity as a function of frequency measured by a T-SPEC 800 terahertz time-domain spectroscopy (THz-TDS) system, revealing pronounced variations among materials [32, 33]. The T-SPEC 800 acquisition setting provided a spectral resolution of 8 GHz. The extracted permittivity spectra were obtained through the THz-TDS extraction procedure, with noise filtering applied before extraction and no additional post-extraction processing. Supplementary Table S-I summarizes the extraction procedure and data processing. For heterogeneous indoor materials, the extracted permittivity represents an effective electromagnetic response arising from the contributions of different constituents and microstructures. Within the investigated frequency range, these effective responses exhibit smooth and gradually varying frequency dependence [18]. Wall plaster exhibits the highest real permittivity ($\varepsilon' \approx 5$), consistent with its mineral-based composition, while curtain shows a lower value ($\varepsilon' \approx$ 1.5-2.0) due to its porous microstructures. Materials with similar base constituents, such as PVC wallpaper and PVC leather, exhibit discernible differences attributable to variations in fabrication processes and filler distributions. A significant contrast is observed between earlywood and latewood specimens, with the latewood region exhibiting consistently higher $\varepsilon'$ and $\varepsilon''$ values due to its greater cell wall density and reduced porosity [34], in agreement with effective medium theory predictions [29, 35]. This density-dependent permittivity variation within a single material leads to spatially varying electromagnetic properties, inducing structure-dependent scattering phenomena that will be examined in the following subsection.

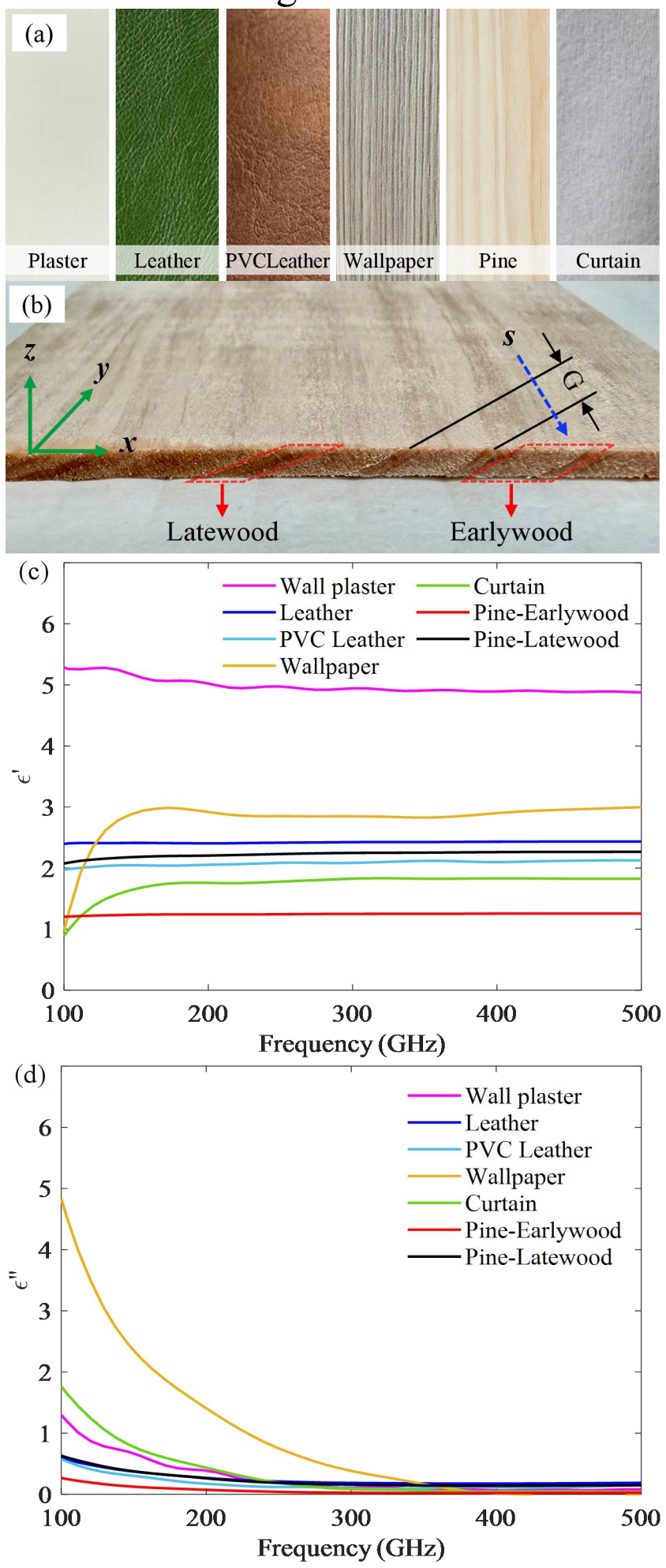

Fig. 1. (a) Surface morphology of the materials under test. (b) Cross-sectional view of the pine wood sample, where $G$ denotes the period of the alternating earlywood-latewood structure, $s$ indicates the orientation of the periodicity, and $x$, $y$, and $z$ represent the geometric axes of the wood sample. (c) Measured real part of the complex relative permittivity of the tested materials as a function of frequency. (d) Measured imaginary part of the complex relative permittivity of the tested materials as a function of frequency. (Thickness of wall plaster: 1.21 mm; leather: 2.22 mm; PVC leather: 1.66 mm; wallpaper: 0.22 mm; pine wood: 2.73 mm for earlywood specimen, 2.03 mm for latewood specimen; and curtain: 0.35 mm).

### B. Experimental Measurement

A continuous-wave (CW) measurement system operating in the D-band was employed, as illustrated in Fig. 2(a). At the transmitter (Tx) side, a signal generator (Ceyear 1465D) produced a stable 0-dBm RF signal, which was up-converted to the 113-173 GHz frequency range using a ×12 frequency multiplier (Ceyear 82406B). Measurements were performed at three representative frequencies, namely 113, 140, and 170 GHz, using the same pair of HD-1400SGAH25 standard-gain horn antennas for both Tx and Rx, with corresponding gains of approximately 24.8, 26.3, and 27.5 dBi, respectively. The E-/H-plane half-power beamwidths are 7°-11° over the operating frequency range. With a 10-cm-focal-length polytetrafluoroethylene (PTFE) lens installed on the Tx horn, the angular response was characterized through free-space measurements, as shown in Fig. S1 of the Supplementary Material. The received power was recorded using a calibrated power sensor (Ceyear 71718). The channel beam was directed onto the material under test (MUT) surface at a controlled incidence angle, with the sample vertically mounted on the optical table. To characterize the angular scattering behavior, the Rx horn was mounted on a computer-controlled rotating platform and scanned with an angular sampling step of 1°. The same rotary stage was used for the reflection and transmission scans. Its angular reference drifted by approximately 0.1° per 100 angular positions, estimated from the scale-mark offset accumulated over several hundred positions. Before each sample measurement, the designated scale mark on the rotary stage was checked for alignment with the fixed index mark to maintain a consistent angular reference, as shown in the inset of Fig. 2(a). During each scan, the Tx assembly remained fixed, and the rotary stage remained stationary at each angle for a 1-s acquisition. Two complete angular scans were averaged in the linear-power domain before conversion to dBm, with no smoothing, angular resampling, curve fitting, or outlier removal applied. For the six materials in Fig. 2(b), the maximum absolute pointwise power differences between scans were 0.30-0.82 dB, with coincident dominant peak angles. These scan-to-scan differences are taken as the power uncertainty of the angular profiles. The angular uncertainty has three sources: the 1° sampling, which locates well-defined lobe maxima to within ±0.5°; the stage drift, limited to below 0.2° per sample measurement by the alignment check; and sample repositioning, checked by the aluminum-plate reference scans below. Shifts of approximately 2° (two steps) are therefore distinguishable, whereas shifts of approximately 1° (one step) lie at the measurement limit. Both MUT-to-Tx and MUT-to-Rx distances were fixed at 34 cm, giving a total path length of 68 cm. To ensure measurement integrity and minimize parasitic reflections, the MUT was carefully positioned relative to surrounding metallic structures, including the support frame's aluminum profiles.

For reflection measurements, a polished aluminum plate served as the total-reflection reference [4]. With the Tx and Rx axes intersecting at 90° at the sample center, the plate orientation was adjusted to maximize the received power and establish the 45° reflection geometry shown in Fig. 2(a). The sample holder was then fixed, with each sample stationary throughout its angular scan. To assess measurement consistency across sample replacements, aluminum-plate reference measurements were performed before and after the reflection measurements of the six materials in Fig. 2(b). Both measurements yielded the same sampled peak angle of 45°, with a difference of 0.045 dB in peak received power. Only 140 GHz results are presented, as observations at 113 and 170 GHz exhibited consistent trends. Fig. 2(b) shows that all tested materials exhibit a dominant specular reflection component centered at the geometric reflection angle. Wall plaster exhibits the strongest reflection while curtain fabric shows the weakest, consistent with the permittivity trend in Fig. 1(c). The reflection results in Fig. 2(b) (red curve) were obtained from a large pine wood plate featuring alternating earlywood and latewood structures (see Fig. 1(b)), thus representing an effective average over the spatially inhomogeneous internal structure [36]. Transmission measurements employed a line-of-sight (LoS) configuration with normal incidence on the material surface, and the receiver rotated to scan angles on both sides of the direct transmission path (0°). Free-space transmission provided the reference baseline. In Fig. 2(b) and (c), most samples exhibit relatively smooth dominant angular envelopes, although local fluctuations remain. The power levels are primarily governed by effective electromagnetic parameters, including complex permittivity and thickness, through interfacial reflection, absorption, and internal reflections, whereas random microstructural inhomogeneities contribute only weak diffuse scattering under the present measurement conditions [5]. In transmission, the quasi-periodic earlywood-latewood structure of pine modulates the complex refractive index, producing more pronounced angular fluctuations and off-axis lobes [35, 37], as analyzed in Section II-C. Fig. S2 of the Supplementary Material provides additional glass and ceiling measurements.

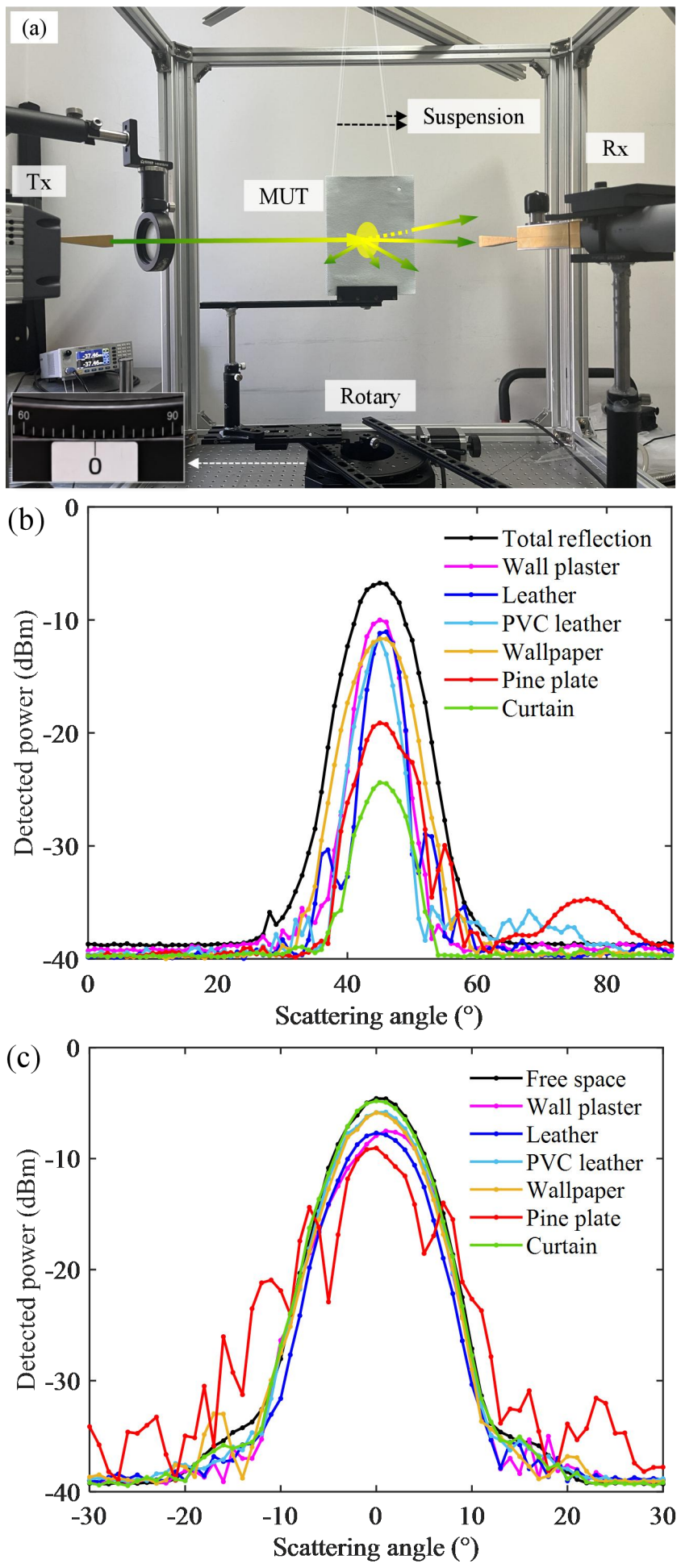

Fig. 2. (a) Terahertz channel measurement setup for angular reflection and scattering characterization. Inset: angular scale and fixed index mark. (b) Measured reflected/scattered power compared with total specular reflection at 140 GHz. (c) Measured transmitted power compared with the free-space reference at 140 GHz.

### *C. Structure-induced Scattering*

Pine wood exhibits quasi-periodic internal inhomogeneity from alternating earlywood (~0.29 g/cm³) and latewood (~0.64 g/cm³) layers [29, 35] resulting in corresponding variations in electromagnetic properties at THz frequencies, as shown in Fig. 1(c)-(d). When cut perpendicular to the growth-ring direction, this internal structure can be regarded as a laterally modulated dielectric medium. As illustrated in Fig. 1(b), the effective spacing between adjacent growth layers is approximately $G$ = 15 mm with a duty cycle of $D$ = 0.25 (the fractional width of the latewood layer within one effective period), and the modulation direction $s$ forms an angle of $\psi$ = 25° with the $z$-axis. These parameters were measured from the cross-sectional region shown in Fig. 1(b). Given the sample geometry (L × W × H: 150 × 100 × 3 mm³) and the effective channel beam diameter of approximately 24 mm, the illuminated growth layers can be reasonably approximated as quasi-parallel stratified structures with uniform inclination, allowing ring curvature to be neglected in subsequent analysis.

To establish a quantitative connection between the experimentally observed angular scattering characteristics and the internal growth-layer structure of pine, the refractive-index inhomogeneity is approximated as a locally periodic tilted modulation over the illuminated aperture, providing a tractable idealization of the quasi-periodic growth-layer morphology. Accordingly, a local coordinate along the growth-layer direction is introduced as

$$s(x,z) = x\sin\psi + z\cos\psi \tag{1}$$

where the modulation direction $s$ and the coordinate system ($x$, $y$, $z$) follow the geometric definitions shown in Fig. 1(b), and the modulation is assumed to be invariant along the $y$-direction. This approximation yields a two-dimensional refractive-index modulation in the $x$-$z$ plane.

Such a structure can be naturally incorporated into a beam-propagation model (BPM) [37] inspired by diffraction theory. The propagation of the complex electric field $E(x, y, z)$ inside the pine sample is governed by the scalar Helmholtz equation [37], as

$$\nabla^2 E + k_0^2 n^2(x,y,z)E = 0 \tag{2}$$

where $k_0$ denotes the free-space wavenumber and $n(x, y, z)$ represents the spatially varying refractive index distribution. To enable efficient numerical propagation, the refractive-index distribution is decomposed into a spatially averaged background component and a position-dependent perturbation

$$n(x,y,z) = n_0 + \delta n(x,y,z) \tag{3}$$

where $n_0$ represents the duty-weighted averaged refractive index of pine, and $\delta n(x, y, z)$ denotes the refractive-index perturbation

$$\delta n(x,y,z) = \begin{cases} \delta n_H, \mathrm{mod}(s,G) < DG \\ \delta n_L, otherwise \end{cases} \tag{4}$$

with $\delta n_H$ and $\delta n_L$ corresponding to the refractive-index deviations of the latewood and earlywood layers, which can be extracted from THz-TDS measurements.

The propagation through the sample thickness $d$ is discretized into $N_z$ steps with step size $\Delta z = d/N_z$. Substituting (3) into (2) and invoking the paraxial approximation, the field evolution is approximated using a split-step beam-propagation scheme, in which transverse diffraction in a homogeneous background medium and the local complex material response are treated sequentially within each propagation step. Following standard beam-propagation and angular-spectrum formulations [38], the diffraction contribution is evaluated in the spatial-frequency domain, yielding the angular-spectrum propagation operator,

$$H(k_x,k_y) = \exp\left(-i\frac{k_x^2+k_y^2}{2k_0 n_0}\Delta z\right) \tag{5}$$

where $k_x$ and $k_y$ denote the transverse wave-vector components. This operator is derived from the angular-spectrum representation of the scalar Helmholtz equation under the paraxial condition $k_x^2 + k_y^2 << (k_0 n_0)^2$, which is satisfied here because the measured and simulated power concentrates within the central lobe and first-order lobes, and the residual power beyond this range is close to the measurement noise floor.

Let the complex refractive index be $n^*(x,y,z) = n(x,y,z)+i\kappa(x,y,z)$, with $n$ the phase index and $\kappa$ the extinction coefficient. We decompose $n(x,y,z) = n_0 + \delta n(x,y,z)$, $\kappa(x,y,z) = \kappa_0 + \delta\kappa(x,y,z)$. The per-step material operator is then

$$M(x,y,z) = \exp(ik_0\delta n(x,y,z)\Delta z)\exp(-k_0\delta\kappa(x,y,z)\Delta z) \tag{6}$$

while the background phase/attenuation due to $n_0$, $\kappa_0$ is included in the homogeneous propagation term.

The BPM calculation used the effective Gaussian input field defined by Eq. (18) in the Appendix, whose width was determined from the measured main-lobe HPBW, so that the angular response of the measurement system is represented through the input field. Reception of the output field by the Rx antenna was not explicitly modeled. The comparison therefore focuses on lobe positions, angular shifts, and main-lobe widths rather than on fine angular structure. After propagation to the exit interface $z = d$, the $k_y = 0$ angular-spectrum cut was extracted for consistency with the receiver scan in the $x$-$z$ plane. Because the transmission measurement is performed at (near-) normal incidence, we neglect refraction-induced remapping of transverse spatial frequencies at the exit interface and map propagating components via $k_x = k_0 \sin\theta_s$ in air. The angular profiles obtained by fast Fourier transform (FFT) were compared with measurements sampled at 1 ° intervals. The detailed numerical implementation of the BPM calculation is summarized in the Appendix. The BPM formulation is not limited to the measured frequencies or the pine sample structure; however, applications to similar structural variations or other scales require updating the wavenumber, frequency-dependent material parameters, internal structural period, tilt angle, duty cycle, refractive-index contrast, and numerical sampling.

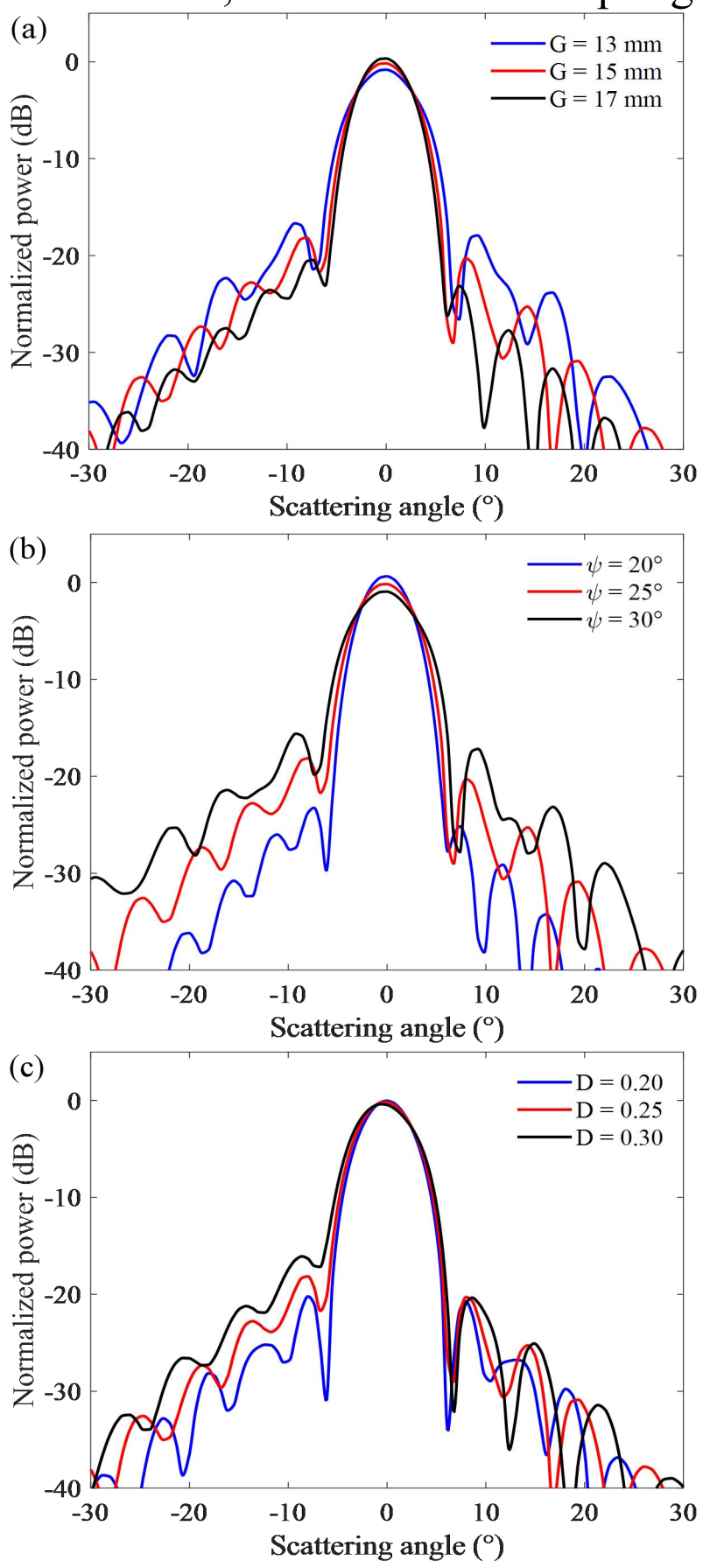

Fig. 3. (a) Calculated angular scattering distributions for different effective periods $G$, where $G$ represents the spacing between adjacent growth layers. (b) Calculated angular scattering distributions for different tilt angles $\psi$, where $\psi$ represents the angle between the modulation direction $s$ and the $z$-axis. (c) Calculated angular scattering distributions for different duty cycles $D$, where $D$ represents the fraction of the latewood layer within one effective period. All distributions were computed at 140 GHz.

Within this modeling framework, the role of key structural parameters in shaping the angular scattering response can be systematically assessed. To further examine the influence of realistic structural variations, the pine plate (see Fig. 1(b)) was translated along the $x$-direction by ±5 mm and ±10 mm relative to the reference position. Since the illuminated region was located near the center of the pine plate, destructive cutting was avoided to preserve the sample. Therefore, the structural parameters used in Fig. 3 were extracted from the corresponding growth-layer features on the incident and exit faces within the illuminated regions over this lateral displacement range. For each identifiable growth-layer period $i$, the effective local period was obtained by averaging the period widths measured on the two faces, i.e., $G_i = (G_{i,incident} + G_{i,exit}) / 2$. The duty cycle $D_i$ was obtained by averaging the latewood-to-period width ratios measured on the incident and exit faces. The tilt angle $\psi_i$ was determined from the lateral displacement $\Delta x_i$ of the same growth-layer reference point between the two faces and the sample thickness $d$, i.e., $\psi_i = \arctan(|\Delta x_i| / d)$. Over the -10 to +10 mm translation range, $G$, $D$, and $\psi$ varied approximately within 13-17 mm, 0.2-0.3, and 20°-30°, respectively. The variation range of $D$ corresponds to an estimated standard deviation of approximately 0.03. These measured ranges were used as the bounds of the representative parameter sweeps in Fig. 3.

Variations in the effective period $G$ constitute one of the key factors governing the angular position of the first side-lobe, as illustrated in Fig. 3(a). By comparing the calculated scattering distributions corresponding to $G$ = 13, 15, and 17 mm, with $\psi$ = 25° and $D$ = 0.25 fixed, an angular shift of approximately 1.8° is observed for the principal off-axis lobe, while the overall envelope of the angular distribution remains largely unchanged. At 140 GHz, the measured first side-lobe in Fig. S3 of the Supplementary Material shifts with position by approximately 2° (two sampling steps), which is consistent, within the ±0.5° quantization of each lobe position, with the calculated shift of 1.8° in Fig. 3(a).

Similarly, the structural tilt angle $\psi$ is found to exert a noticeable influence on the angular scattering characteristics, as illustrated in Fig. 3(b). Specifically, $G$ and $D$ are fixed at 15 mm and 0.25, respectively, while $\psi$ is varied among 20°, 25°, and 30°. As $\psi$ increases, the side-lobes gradually shift further away, with the first side-lobe shifting by approximately 1.8°. This behavior reflects the role of $\psi$ in modifying the effective transverse projection of the internal periodic structure, as defined in (1), which governs the transverse coupling efficiency and consequently shapes the angular redistribution of scattered energy [37].

In contrast to $G$ and $\psi$, variations in the duty cycle $D$ exert a relatively minor influence on the angular position of the first side-lobe, producing a shift of about 0.6°, while only slightly modifying the overall scattering envelope. As shown in Fig. 3(c), the calculated main-lobe shift is about 0.6°, and the measured main-lobe position in Fig. S3 varies by approximately 1° (one sampling step). Both values lie at the respective sampling limits. They indicate only that the main-lobe direction is insensitive to D within the examined range and are not used as quantitative validation. Here, $G$ and $\psi$ are fixed at 15 mm and 25°, while $D$ is varied among 0.2, 0.25, and 0.3. This behavior indicates that the duty cycle governs the amplitude of the internal refractive-index modulation, thereby regulating the transverse coupling efficiency into different angular scattering channels without introducing additional angular components [39]. The corresponding angular-feature values are summarized in Table II.

The extracted structural parameters may vary with natural growth-related density variations and the relative orientation between the artificial cutting direction and the annual-ring structure. To evaluate their joint influence beyond independent parameter sweeps, the 3×3×3 parameter grid of $G$, $\psi$, and $D$ was further analyzed by comparing first-side-lobe angular-position ranges. The extracted first-side-lobe angular positions and detailed comparisons are provided in Supplementary Table S-II. Across the grid, the main-lobe direction varies over an angular span of approximately 0.6°, indicating that the dominant propagation direction remains comparatively robust within the examined parameter ranges. All three parameters affect the selected first-side-lobe position, but their conditional influences differ across the parameter grid. The angular range induced by varying $G$ changes with $\psi$ at every investigated $D$ level, indicating the $G$-$\psi$ pair exhibits the most consistently observed condition-dependent response among the examined parameter pairs. $D$ also participates in the joint response; under the adopted angular-range metric, its induced range remains within 0.6°-0.7° in eight of the nine fixed ($G$, $\psi$) settings and reaches 1.8° only at $G$ = 17 mm and $\psi$ = 20°. This $G$-$\psi$ dependence is consistent with the respective roles of $G$ and $\psi$ in determining the magnitude and orientation of the spatial-frequency vector associated with the tilted refractive-index modulation [37, 39].

TABLE II

QUANTITATIVE ANGULAR-FEATURE COMPARISON FOR THE PINE SAMPLE

| Case | First side lobe shift | Main-lobe shift | Main-lobe width $W_{3dB}$ | Mean Pearson coefficient $r_{mean}$ |
|---|---|---|---|---|
| $G$ variation | 1.8° | 0° | 5.2° - 6.4° | 0.843 |
| $\psi$ variation | 1.8° | 0° | 4.8° - 6.5° | 0.833 |
| $D$ variation | 0.6° | 0.6° | 5.2° - 6.3° | 0.828 |
| Measurement | 2° | 1° | 3.2° - 7.5° | - |

For comparison with the measured main-lobe broadening, the -3 dB main-lobe widths $W_{3dB}$ were extracted from the calculated angular responses under different structural parameters and from the measured responses at different translated positions. The width was determined by locating the two angular points at which the response falls 3 dB below the central main-lobe peak and taking their angular separation. The extracted widths range from 4.8° to 6.5° for the BPM predictions and from 3.2° to 7.5° for the measurements, as shown in Table II. Although the measured range is slightly broader, the calculated and measured widths remain broadly comparable, with differences of approximately 1.6° and 1° between their lower and upper bounds, respectively. Together with the comparisons of the first side-lobe shift and the main-lobe angular shift, the discrepancies between the model predictions and measurements in the main angular features are generally on the order of 1° - 2°. This indicates that, despite structural inhomogeneity and measurement resolution limits, the BPM model can reasonably capture the dominant angular scattering features induced by variations in the internal periodic structure of the pine sample.

In addition, the Pearson correlation coefficient $r$ was used as a similarity metric to evaluate the overall profile agreement between the measured angular responses and the BPM-predicted angular responses over the angular range covering the central main lobe and the first off-axis lobe. It is defined as

$$r = \frac{\sum_j \left(P_{meas,j} - \overline{P}_{meas}\right)\left(P_{model,j} - \overline{P}_{model}\right)}{\sqrt{\sum_j \left(P_{meas,j} - \overline{P}_{meas}\right)^2}\sqrt{\sum_i \left(P_{model,j} - \overline{P}_{model}\right)^2}} \quad (7)$$

where $P_{meas,j}$ and $P_{model,j}$ denote the peak-normalized measured and BPM-predicted angular responses at the $j$-th angular sampling point, respectively. $\overline{P}_{meas}$ and $\overline{P}_{model}$ are the corresponding mean values over the selected angular range. For each parameter sweep, the BPM response computed at the structural parameters extracted for a given translated position was correlated with the measured response at that position over the angular window spanning the central main lobe and the first off-axis lobe, and $r_{mean}$ denotes the average over the positions. The resulting values are 0.843, 0.833, and 0.828 for the $G$, $\psi$, and $D$ sweeps, as shown in Table II. Because this window is dominated by the main lobe, $r_{mean}$ principally reflects main-lobe agreement and is reported as a complement to - not a substitute for - the first-side-lobe shift, which is the more discriminating feature. These results indicate that the BPM model can reasonably capture the main scattering features and the overall angular-response profile induced by variations in the internal periodic structure of the pine sample.

Before extending the analysis to covered surfaces and indoor environments, we state the evaluation rationale applied uniformly in the remainder of this work. Each quantitative comparison is of one of two kinds, and the distinction is made per comparison rather than per section. Where a predictive model is available, agreement is reported as a model-versus-measurement comparison, using the metric matched to the dominant feature of the angular response: the first-side-lobe and main-lobe angular shifts for the discrete-lobe pine response, complemented by the Pearson correlation $r$ as an overall profile-similarity measure; the frequency-trend Pearson $r$ for the frequency-selective suppression of the smooth-covered surface (Section III-A); and the root-mean-square (RMS) angular spread $\sigma$ together with the normalized mean-square error $NMSE_{dB}$ for the distributed folded-curtain response (Section IV), which characterize its angular scale and pointwise shape, respectively. Where no predictive model is introduced, as for the rough-covered surfaces (Section III-B), the same class of physical observables - peak power, peak angle, main-lobe width, and integrated profile power - is reported relative to a measured bare-substrate reference under identical geometry. A single section may contain both kinds of comparison; in Section III-A, the specular-power reduction relative to the bare aluminum plate is a measurement-versus-reference change, whereas the thin-film model is validated against the measured frequency trend as a model-versus-measurement comparison. The metric in each case is thus determined by whether a model is validated or measured response is being compared to a baseline, and by which feature of the angular response carries the physical information.

## III. CHARACTERIZATION OF COVERED SURFACES

Building upon the structure-oriented scattering mechanisms identified for single-material interfaces in Section II, this section extends the analysis to composite boundary conditions formed when a substrate is covered by an additional material layer. Such material-covered surfaces are ubiquitous in practical indoor environments, commonly encountered when metallic frames, cabinets, appliances, and structural elements are partially or fully coated by wallpapers, fabrics, decorative laminates, or other functional coverings.

### *A. Scattering by Covered Smooth Surfaces*

We consider a practically relevant configuration consisting of a smooth metallic reflector covered by a homogeneous dielectric material layer. This scenario commonly arises in indoor environments when metallic panels, frames, or appliances are partially or fully covered by decorative or functional materials. By employing a polished aluminum plate as the reference substrate, geometric irregularities and surface-roughness-induced distortions are minimized, thereby allowing the analysis to focus specifically on the electromagnetic effects introduced by the material covering layer on the reflected power and angular response.

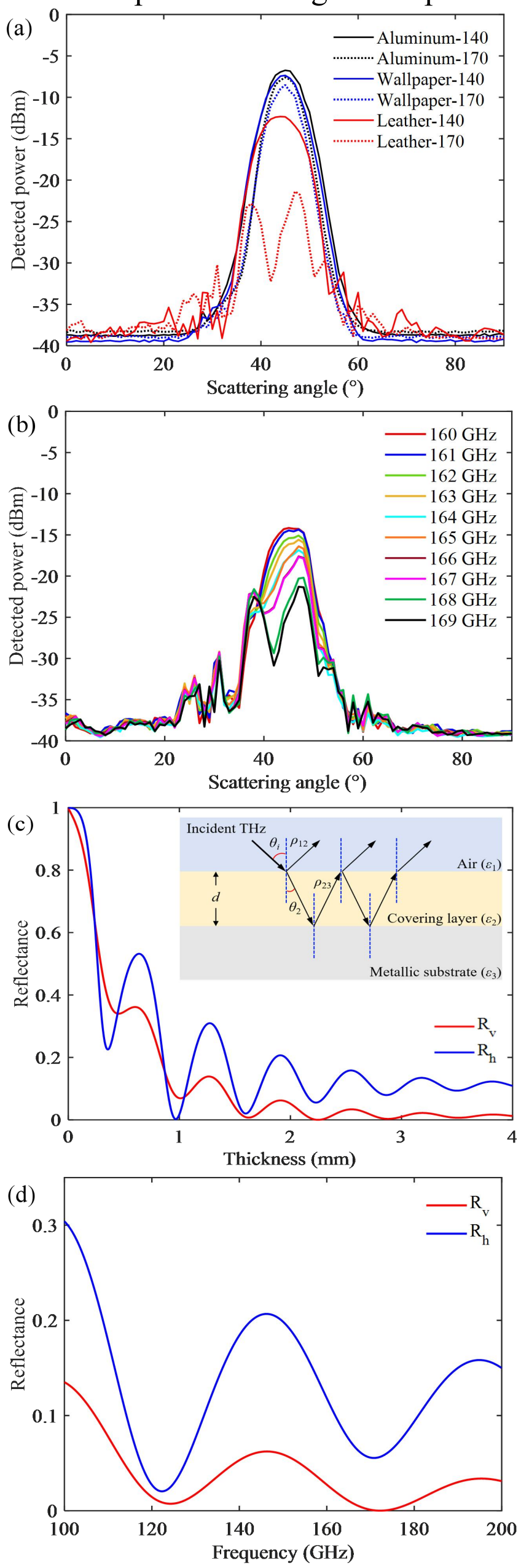

Fig. 4. (a) Scattering distributions of material-covered surface. (b) Frequency-dependent scattering distributions of an aluminum plate covered by leather. (c) Calculated reflectance at 170 GHz as a function of the covering material thickness; the inset illustrates the three-layer reflection model with internal reflections. (d) Calculated reflectance as a function of frequency for a fixed covering thickness.

Under the same reflection geometry described in Section II-B and illustrated in Fig. 2(a), angular scattering distributions were measured for a bare aluminum plate and for the same plate covered with representative indoor materials, namely wallpaper and leather. The aluminum plate has dimensions of 100 mm × 100 mm × 5 mm. Each covering layer was cut to identical dimensions and attached along the edges to ensure near-conformal contact with the metallic substrate, thereby minimizing residual air gaps between the covering layer and the metal surface.

Angular scattering distributions were measured at 113, 140, and 170 GHz. For clarity, Fig. 4(a) presents representative results at 140 and 170 GHz, highlighting the contrast between typical scattering behavior and the frequency-dependent deviations observed at higher frequencies. Compared with the bare aluminum reference, the material-covered surfaces exhibit reduced detected power near the specular direction (45°). The relative reductions are approximately 0.6 dB and 5.6 dB for wallpaper and leather at 140 GHz, respectively, and become 0.9 dB and 17.5 dB at 170 GHz. These reductions reflect additional electromagnetic effects introduced by the covering layers, such as loss, impedance mismatch, and frequency-dependent interference effects. For the wallpaper-covered configuration, the angular profile remains similar to that of the bare aluminum plate, suggesting that the response continues to be dominated by specular reflection from the underlying metal surface, with the thin wallpaper layer contributing primarily to additional insertion loss rather than altering the angular shape of the specular lobe.

In contrast, the leather-covered configuration at 170 GHz exhibits a pronounced suppression and deformation of the angular response near the specular direction, deviating substantially from the expected single-peaked profile. To investigate whether this anomalous behavior exhibits frequency dependence, additional angular scattering measurements were conducted for the leather-covered surface from 160 to 169 GHz in 1 GHz increments, as shown in Fig. 4(b). As the frequency decreases from 169 to 160 GHz, the distorted profile gradually evolves into a well-defined single-peaked specular response, and the detected power near the specular direction increases by approximately 10.9 dB relative to that at 170 GHz. This trend confirms that the distortion observed at 170 GHz is strongly frequency dependent. Given that the leather layer (thickness 2.22 mm) is substantially thicker than the other covering materials examined, the observed behavior is hypothesized to be associated with the effective optical thickness of the covering layer. Specifically, coherent multiple reflections within the dielectric layer can substantially modify the effective reflectance at the metal interface through thin-film interference effects [40].

To quantitatively examine this hypothesis, the material-covered metallic surface is modeled as a planar three-layer structure consisting of air (medium 1), a lossy dielectric covering layer (medium 2), and a metallic substrate (medium 3), as schematically shown in the inset of Fig. 4(c). The analysis focuses on the coherent specular reflection component and adopts a transfer-matrix thin-film formulation in which multiple internal reflections and material absorption are explicitly incorporated. For an obliquely incident plane wave at angle $\theta_i$ with respect to the surface normal, refraction inside the dielectric layer is governed by Snell's law generalized to complex permittivity. Assuming non-magnetic media ($\mu_r \approx 1$) [41, 42], the complex refraction angle inside the covering layer can be expressed [9] as

$$\theta_2 = \arcsin\left(\frac{\sqrt{\varepsilon_1}}{\sqrt{\varepsilon_2}}\sin\theta_i\right) \tag{8}$$

where $\varepsilon_1$, $\varepsilon_2$, $\varepsilon_3$ denote the complex permittivities of air, the covering material, and the metallic substrate, respectively. The Fresnel reflection coefficients at the air-cover interface ( $\rho_{h/v}^{12}$ ) and the cover-metal interface ( $\rho_{h/v}^{23}$ ) are computed for both horizontal ($h$) and vertical ($v$) polarizations using the standard Fresnel equations. The coherent round-trip propagation inside the covering layer of thickness $d$ is described by the complex phase factor [43]

$$P(d,f) = \exp(-2\alpha d\cos\theta_2)\exp(-j2\beta d\cos\theta_2) \tag{9}$$

where $f$ is the operating frequency. The phase constant $\beta$ and attenuation constant $\alpha$ inside the lossy dielectric layer are implicitly contained within the complex propagation constant $k_2 = k_0\sqrt{\varepsilon_2} = \beta - j\alpha$ . The longitudinal propagation constant inside the covering layer is given by $k_{z,2} = k_2\cos\theta_2$. Since the covering material exhibits non-negligible loss, $\varepsilon_2$ is complex-valued as determined by the THz-TDS measurements presented in Section II-A, and (9) inherently accounts for both phase accumulation and absorption-induced amplitude decay during coherent round-trip propagation within the layer.

The coherent specular reflection from the material-covered metallic interface is modeled using the Fresnel-Airy formulation for a lossy dielectric slab. The effective complex reflection coefficient is obtained by summing the multiple internal reflections within the covering layer, as

$$\rho_{h/v}(d,f) = \frac{\rho_{h/v}^{12} + \rho_{h/v}^{23}P(d,f)}{1+\rho_{h/v}^{12}\rho_{h/v}^{23}P(d,f)} \tag{10}$$

and the corresponding specular reflectance is given by

$$R_{h/v}(d,f) = \left|\rho_{h/v}(d,f)\right|^2 \tag{11}$$

Based on the above thin-film interference model, Fig. 4(c) presents the calculated reflectance at 170 GHz as a function of covering layer thickness for both horizontal ($R_h$) and vertical ($R_v$) polarizations, using the complex permittivity values extracted from THz-TDS measurements. As the thickness varies, pronounced oscillations in both $R_h$ and $R_v$ are observed, arising from the combined effects of coherent phase interference and material absorption within the covering layer [44]. Notably, for a leather thickness of 2.22 mm corresponding to the actual sample, the calculated reflectance reaches a local minimum for both polarizations, in good agreement with the experimentally observed power suppression at 170 GHz shown in Fig. 4(a).

TABLE III
QUANTITATIVE COMPARISON OF SPECULAR-DIRECTION RESPONSES FOR COVERED SMOOTH SURFACES

| Metric | Frequency | Wallpaper-covered | Leather-covered |
|---|---|---|---|
| Power reduction from bare Al plate | 140 GHz | 0.6 dB | 5.6 dB |
| | 170 GHz | 0.9 dB | 17.5 dB |
| Pearson correlation coefficient $r_{h/v}$ | 160-170 GHz | - | $r_h = 0.948$<br>$r_v = 0.984$ |

Fig. 4(d) further presents the calculated reflectance as a function of frequency for the fixed leather covering thickness of 2.22 mm. As the frequency increases from 100 GHz toward 170 GHz, the reflectance exhibits an oscillatory frequency-dependent variation and reaches a pronounced minimum near 170 GHz, while remaining significantly higher at frequencies around 140 GHz. This frequency-dependent trend closely reproduces the evolution observed in the experimental measurements of Fig. 4(b). To quantify the agreement in the relative frequency-dependent trend, the measured specular-power data at 45° and the calculated reflectance spectra over 160-170 GHz were converted to the dB scale and referenced to their respective values at 170 GHz. The resulting Pearson correlation coefficients are $r_h = 0.948$ and $r_v = 0.984$ for $R_h$ and $R_v$, respectively, as summarized in Table III. This quantitative consistency further supports that the anomalous scattering behavior arises from coherent thin-film interference rather than surface roughness or other geometric effects. These results demonstrate that even for a geometrically smooth metallic reflector, a lossy dielectric covering layer can substantially modify the angular scattering response through thickness- and frequency-dependent interference effects. Such phenomena have significant implications for terahertz channel modeling in indoor environments, where material-covered metallic surfaces are commonly encountered and may exhibit strongly frequency-selective reflection characteristics that deviate markedly from simple Fresnel predictions. Extension to other material-covered surfaces requires re-evaluating the layer thickness, dielectric properties of the covering layer and substrate, and interfacial conditions.

### *B. Scattering by Covered Rough Surfaces*

Practical indoor environments frequently contain surfaces with non-negligible surface roughness arising from manufacturing processes, surface wear, or environmental degradation. To address this more general scenario, we investigate the combined influence of material coverings and substrate surface roughness on THz channel scattering behavior. Wallpaper and leather samples were applied to an aluminum plate with controlled surface roughness, characterized by a RMS height of approximately 0.2 mm, as detailed in Fig. A1(*f*) in [45]. This roughness scale is comparable to the operating wavelength in the D-band frequency range, placing the surface in the intermediate roughness regime where both coherent (specular) and incoherent (diffuse) scattering components contribute significantly to the total reflected field [9, 13]. Measurements were performed at 113, 140, and 170 GHz under the same angular-scanning configuration described above.

As illustrated in Fig. 5, the scattering response of the material-covered rough aluminum surface exhibits distinct characteristics depending on the covering layer. To identify the physical mechanism governing the covered rough-surface response, the results obtained from the smooth substrate in Fig. 4(a) are quantitatively compared with those from the rough substrate in Fig. 5(a)-(c). Table IV summarizes the quantitative changes relative to the bare rough aluminum (Al) surface. Consistent with the evaluation rationale of Section II-C, and because this subsection introduces no independent predictive model, these quantities are reported as measurement-versus-reference changes relative to the bare rough-aluminum surface rather than as model-versus-measurement errors. With the wallpaper covering, the specular response of the rough substrate is strengthened, with the peak power increasing by approximately 1.2, 1.4, and 3.9 dB at 113, 140, and 170 GHz, respectively. In contrast, the smooth substrate exhibits an overall attenuation trend after wallpaper covering. Moreover, the corresponding peak angle remains within 1° of that of the bare rough aluminum surface at all three frequencies, indicating that the enhancement occurs without noticeably shifting the dominant reflection direction. In addition, off-specular scattering is notably suppressed, particularly within 10°-30°. In terms of beamwidth, the extracted -3 dB main-lobe width changes only moderately by approximately 0.3°-2.4° after wallpaper covering. Taken together, the wallpaper-covered response is more concentrated around the expected reflection direction and approaches an idealized specular profile. These comparisons indicate that the primary role of the wallpaper covering on the rough substrate is to effectively smooth the rough electromagnetic interface. Specifically, the thin dielectric layer induces a redistribution of the angular scattering spectrum, causing part of the scattering energy originally distributed into off-specular directions to be redirected toward the specular direction. As a result, the coherent reflection component is enhanced, while the diffuse scattering from the underlying rough metallic surface is suppressed. This phenomenon is analogous to the well-documented effect of dielectric coatings in reducing radar cross-section fluctuations from rough conducting surfaces, where the covering layer serves to partially smooth the effective electromagnetic interface [46, 47].

In comparison, when leather is applied to the smooth substrate, the peak power decreases by 5.5 dB and 14.7 dB at 140 and 170 GHz, respectively (peak of the angular profile; the reductions at 45° in Section III-A are 5.6 dB and 17.5 dB). For the rough substrate, however, the corresponding reductions are only 3.0 dB and 4.0 dB at 140 and 170 GHz. At 113 GHz, the leather-covered rough substrate exhibits a notably stronger specular-peak reduction (-11.6 dB) than these 3-4 dB reductions, while the integrated profile power decreases less steeply (-9.5 dB). This suggests that although the leather covering suppresses roughness-induced diffuse scattering to some extent, its relatively large thickness, stronger absorption, and more complex interfacial conformality still weaken the specular reflection component. Therefore, the effect of the leather covering on the rough substrate is more strongly influenced by thickness-related attenuation and nonuniform interfacial geometric conformality, which is also an important reason for the weakened coherent thin-film-interference behavior at 170 GHz. In addition, the relatively large peak-angle shifts of approximately 5°-6° further support the interpretation that the surface-geometry modification introduced by the leather covering is an important contributing factor, including local contact conditions, local thickness variations, and local tilt-angle changes. However, with the present sample set, thickness-induced attenuation, intrinsic absorption, and interfacial conformality cannot be fully decoupled; the dominant role of geometric-conformality effects in the leather-covered rough-substrate case is inferred from the 5°-6° peak-angle shift, which is a characteristic fingerprint of local-tilt variation, rather than independently isolated.

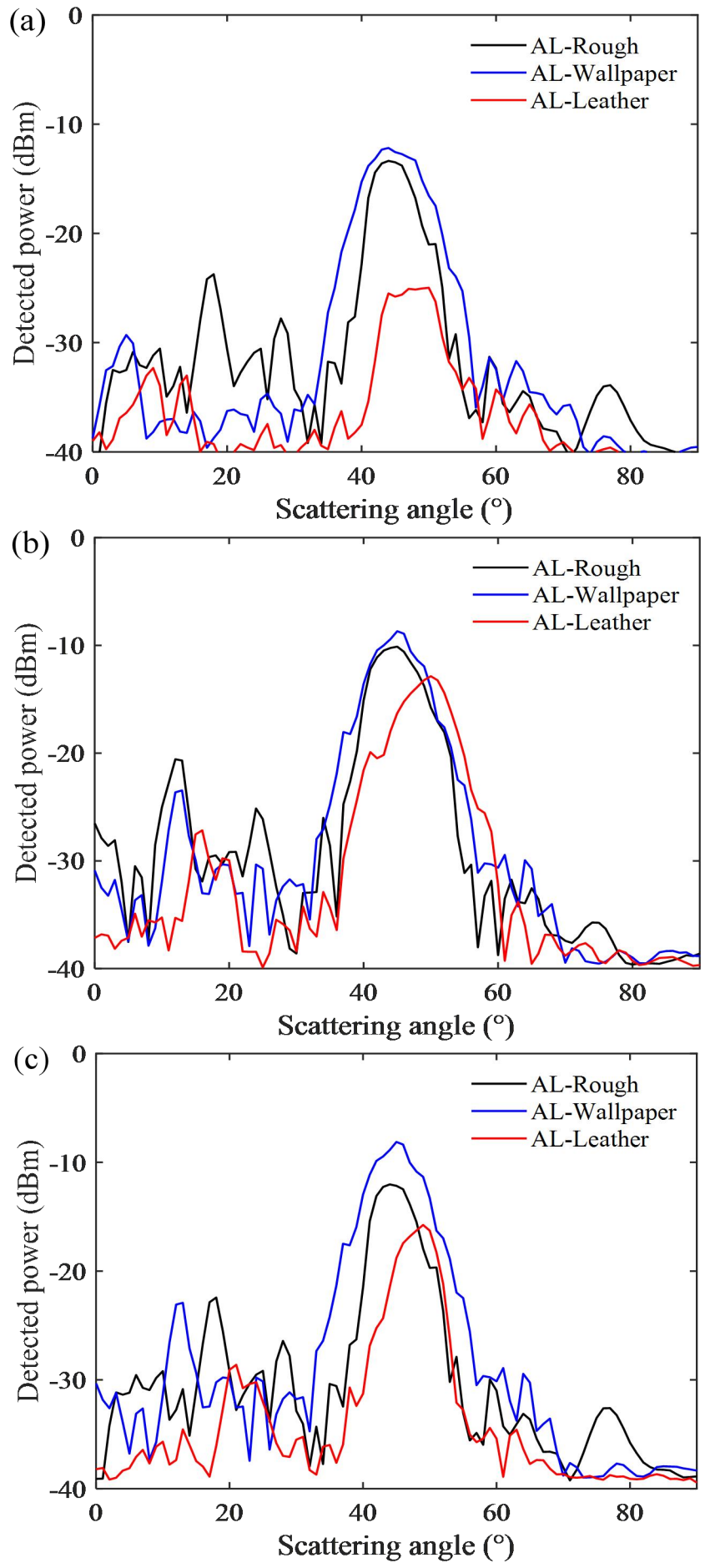

Fig. 5. Scattering distributions of a rough aluminum surface without and with material coverings: (a) 113 GHz; (b) 140 GHz; (c) 170 GHz.

To provide a profile-level comparison, the relative change in the total power of the angular profile within the observed angular range $\Omega_\theta$ = 0°-90° was calculated as follows [51]:

$$\Delta P_{\Omega,c} = 10\log_{10}\left(\frac{\sum_{\theta_s\in\Omega_\theta} p_c(\theta_s)}{\sum_{\theta_s\in\Omega_\theta} p_b(\theta_s)}\right) \qquad (12)$$

where $b$ denotes the bare rough aluminum reference, $c$ denotes the corresponding covered rough surface, $p_b(\theta_s)$ and $p_c(\theta_s)$ are the angular power profiles in linear units, and $\theta_s$ is the scattering angle.

For the wallpaper-covered rough surface, the positive $\Delta P_{\Omega,c}$ values of 2.1, 1.1, and 4.1 dB at 113, 140, and 170 GHz, respectively, indicate an overall enhancement of the measured angular profile. Together with the peak enhancement and off-specular suppression, this suggests scattering redistribution toward the near-specular direction. In contrast, the negative values for the leather-covered case, -9.5, -2.8, and -4.2 dB, indicate an attenuation-dominated profile response, consistent with the combined effects of absorption and nonuniform interfacial conformality.

TABLE IV
QUANTITATIVE COMPARISON OF MEASURED ANGULAR-RESPONSE FEATURES FOR COVERED ROUGH SURFACES

| Covering layer | Wallpaper | | | Leather | | |
|---|---|---|---|---|---|---|
| Frequency (GHz) | 113 | 140 | 170 | 113 | 140 | 170 |
| Peak-power change relative to bare Al (dB) | 1.2 | 1.4 | 3.9 | -11.6 | -3.0 | -4.0 |
| Peak-angle shift relative to bare Al (°) | 0 | 0 | 1 | 6 | 5 | 5 |
| -3 dB main-lobe-width change relative to bare Al (°) | 2.4 | -0.3 | 1 | 2.1 | -0.4 | -0.4 |
| Total angular-profile power change relative to bare Al (dB) | 2.1 | 1.1 | 4.1 | -9.5 | -2.8 | -4.2 |

The above analysis provides a quantitative comparison of the covered-surface responses. By jointly considering the reflection characteristics observed for single-material samples (Fig. 2), smooth covered surfaces (Fig. 4), and rough covered surfaces (Fig. 5), the results show that the scattering response of a covered surface cannot be explained solely by the intrinsic dielectric properties of the covering material. Instead, it is jointly governed by the covering-layer optical thickness, coherent multiple-reflection effects, substrate roughness, and the effective contact condition between the covering and substrate. These factors represent the main material and structural uncertainty sources in covered-surface configurations, because they govern phase accumulation, absorption, and the balance between coherent specular reflection and diffuse angular redistribution. This underscores the importance of considering composite boundary conditions in terahertz channel modeling for realistic indoor environments, where material-covered surfaces with varying substrate roughness are commonly encountered.

## IV. INDOOR ENVIRONMENT-AWARE SCATTERING

Unlike the controlled surface configurations examined previously, indoor scenarios introduce complex propagation effects jointly shaped by furniture arrangements, room geometry, wall and window compositions, and spatially structured reflection and scattering paths [5, 18, 48]. To characterize these environment-level phenomena, wide-angle bistatic measurements were conducted in a typical indoor conference-room setting.

Indoor channel measurements were conducted at two representative frequencies, 140 GHz and 220 GHz, for THz communications. In addition to the D-band experimental configuration described in Section II-B, an ×18 frequency multiplier module (Ceyear 82406D) was employed together with corresponding horn antennas (Ceyear 89901S) and a calibrated power sensor (Ceyear 87106B), enabling terahertz signal generation and detection in the 220-325 GHz frequency band. For both frequency bands, a dielectric lens with a focal length of 35 cm was mounted in front of the horn antenna to collimate the radiated beam, thereby mitigating beam divergence and reducing free-space propagation loss over the extended indoor measurement distances. The measurements were performed in a conference room with dimensions of 7.69 m × 6.93 m, representative of typical indoor meeting spaces encountered in office environments. As illustrated in Fig. 6, the transmitter (Tx) and receiver (Rx) were positioned on the east and west sides of the room with a separation of 3.8 m. Both Tx and Rx were mounted on independent motorized rotary stages for full 360° azimuthal scanning. Both antennas were fixed at a height of 1.15 m above the floor, exceeding the height of window sills and major furniture, to reduce unintended blockage, maintain consistent line-of-sight conditions, and define a controlled horizontal angular measurement plane across different Tx/Rx orientations.

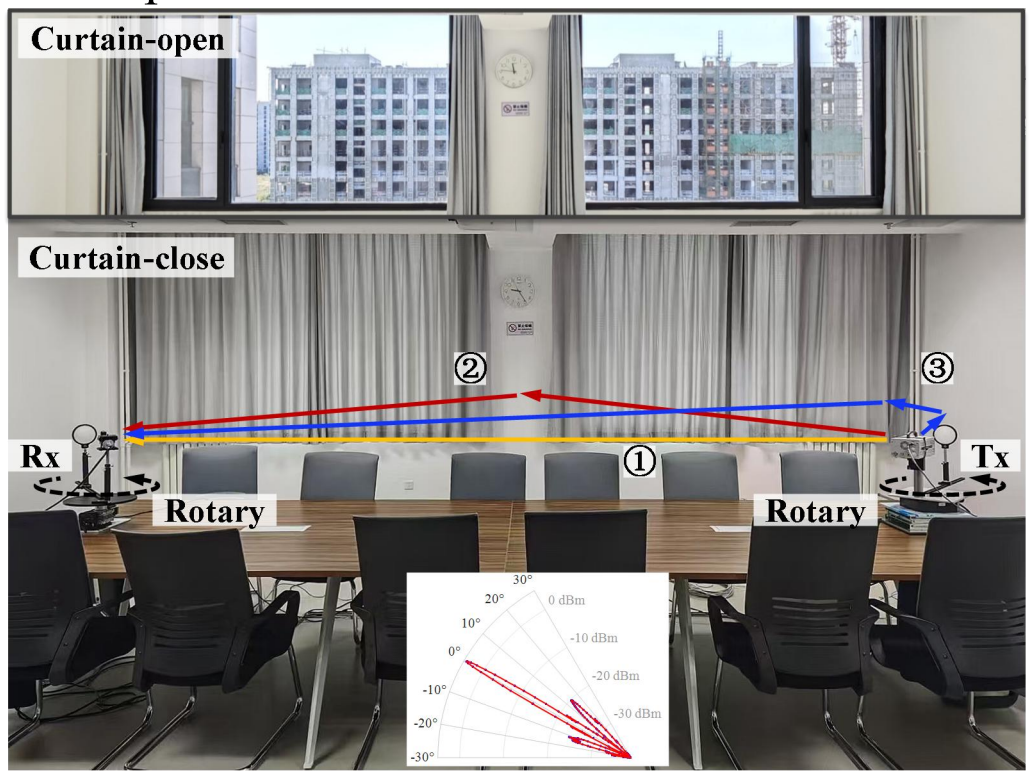

Fig. 6. Indoor terahertz channel measurement setup in a conference-room environment. Top inset: The conference room with window curtains fully opened (Curtain-open). Bottom inset: Effective radiation pattern of the antenna-lens setup at 140 GHz.

The room geometry and material composition are critical factors governing the indoor scattering environment. The south side of the room consists of glass windows and adjacent wall sections. A central wall segment with an east-west width of approximately 0.4 m is positioned between two glass windows and serves as one of the dominant specular reflection surfaces. The glass windows on both sides have widths of 2.845 m and are equipped with fabric curtains that can be either opened or closed depending on the measurement scenario. The configuration illustrated in Fig. 6, labeled "Curtain-close", corresponds to the scenario where the curtains fully cover the glass window surfaces, introducing folded-fabric scattering effects analyzed in Section IV and further supported in the Supplementary Material. On the north side of the room, two doorways (approximately 2.1 m wide each) are located near the corners, and a recessed bar counter (approximately 0.6 m deep and 3.53 m in east-west width) is situated at the center, forming another prominent specular-reflection region.

To achieve comprehensive angular characterization while maintaining practical measurement efficiency, the transmitter was operated with an angular step size of 2°. For each fixed Tx orientation, the receiver rotated continuously at an angular speed of 2°/s, while the received power was recorded at a sampling rate of 7 Hz. The scanning angular range for both Tx and Rx was defined with respect to the line-of-sight direction and spanned ±135°, covering a total azimuthal range of 270°. The positive angular direction corresponds to rotation toward the window (south) side of the room, whereas the negative direction corresponds to rotation toward the bar-counter (north) side. In total, 801826 measurement samples were collected, each consisting of the transmit steering angle $\theta$, the receive observation angle $\varphi$, and the measured received power $P(\theta, \varphi)$. A received-power threshold of -39 dBm is used to identify detectable angular samples in Fig. 7(a)-(d).

The bottom inset of Fig. 6 presents the effective radiation pattern of the antenna-lens assembly, which serves as a reference for interpreting the subsequent indoor scattering measurements. The transmitted beam exhibits strong directivity with a narrow main lobe, indicating a highly concentrated angular power distribution characteristic of terahertz systems. In addition to the main lobe, two sidelobes are observed at approximately ±13° from the boresight direction. These sidelobe components, although substantially weaker than the main beam, can illuminate reflecting surfaces at off-axis angles and thereby give rise to additional detectable propagation paths beyond the dominant specular direction.

Fig. 7(a) presents the full-angle scattering distribution measured at 140 GHz in the conference room with window curtains fully opened, corresponding to the "Curtain-open" configuration illustrated in the top inset of Fig. 6. When both $\theta > 0°$ and $\varphi > 0°$, corresponding to orientations toward the window-facing (south) side of the room, a dominant high-power scattering region is observed around 61°. Based on geometric analysis of the room layout, these components are attributed to specular reflections from the central wall segment (approximately 0.4 m in width) positioned between the two windows. Similarly, for $\theta < 0°$ and $\varphi < 0°$, another high-power scattering region appears around -61°, corresponding to reflections from the bar-counter wall on the north side. Notably, the south-side scattering region extends over a broader angular range than the north side, attributable to the combined effects of partially visible curtain edges, aluminum window frames, and local structural discontinuities that jointly enhance angular spreading in the window-facing region.

To examine how the introduction of a spatially structured element modifies indoor propagation characteristics - building upon the structure-induced scattering mechanisms established in Sections II-C and III - the curtains were deployed to fully cover the windows, corresponding to the "Curtain-close" configuration. The analysis focuses on the angular region where $\theta > 0°$ and $\varphi > 0°$. As shown in Fig. 7(b), a dense set of additional detectable propagation components emerges beyond the original specular-reflection region associated with the central wall segment, extending toward the curtain-covered window areas. To quantify this extension, detectable samples in the window-side region were counted in 2° AoA intervals for both configurations, and the positive count differences (Curtain-close minus Curtain-open) were summed. Approximately 87.5% of this increase lies within ±35° of the original specular direction (≈61°). The remainder occurs at larger offsets with lower density. This count-based measure is analogous to the power-based angular window of [51]. It describes the angular extent of the added detectable samples, not a communication coverage area.

Importantly, the dominant central-wall specular component is preserved at a level comparable to the "Curtain-open" case (Fig. 7(a)), indicating that the curtain-induced scattering predominantly adds propagation paths rather than redistributing the existing specular energy. These additional components arise from the combination of the curtain's intrinsic reflective properties (characterized in Fig. 2(b)) and its folded geometric structure, which together introduce diffuse scattering mechanisms consistent with the material-covered surface analysis in Section III.

In regions where $\theta \cdot \varphi < 0$ (i.e., transmitter and receiver oriented toward opposite sides of the room), detected signal components are generally weak and exhibit spatially irregular distributions. These components represent the ensemble of observable propagation contributions jointly shaped by material properties, environmental geometry, and measurement system characteristics, primarily originating from higher-order reflections, diffuse scattering, sidelobe-assisted propagation paths, and partially captured specular reflections due to finite receiver angular coverage.

To isolate frequency-dependent effects, additional measurements were performed at 220 GHz for both "Curtain-open" and "Curtain-close" configurations, with the angular scan restricted to the region where $\theta \cdot \varphi > 0$, as shown in Fig. 7(c) and Fig. 7(d). This region is dominated by specular components, making curtain-induced variations the primary observable factor. As anticipated, curtain deployment significantly enhances the angular scattering response. Due to the increased directivity at 220 GHz, detectable components in the region where $\theta < 0°$ and $\varphi < 0°$ remain nearly unchanged in both spatial extent and power level. It should be noted that in the "Curtain-close" configuration, a portion of curtain-related scattered energy may not be fully captured due to finite angular coverage, implying that the measured scattering distributions represent conservative estimates of actual channel richness.

To interpret the curtain-induced angular scattering enhancement, a modeling framework is developed by combining local Fresnel reflection for layered media with physical-optics (PO) coherent integration [49]. It focuses on the "Curtain-close" configuration, where the folded curtain forms a spatially structured surface producing additional scattering paths beyond those from the bare glass window. The framework captures the relative angular redistribution of reflected power and can be re-parameterized for other scenarios by updating the relevant geometric configuration. Consider a plane wave with frequency $f$ and incident angle $\theta_i$ impinging on a curtain in front of a glass window. The curtain surface profile is modeled as a one-dimensional (1D) cosine function along the horizontal coordinate $y$:

$$z(y) = A_c \cos\left(\frac{2\pi y}{G_c}\right) \tag{13}$$

where $A_c$ and $G_c$ denote the amplitude and spatial period of the curtain folds. The local surface tilt angle is given by:

$$\alpha(y) = \tan^{-1}\left(\frac{dz}{dy}\right) \tag{14}$$

which determines the local incident angle $\theta_{\mathrm{loc}}(y)$ between the incident wave vector and the local outward surface normal. The electromagnetic response of each illuminated surface element is represented by a local reflection weight. For the practical window configuration, the structure is modeled as a multilayer stack consisting of air, curtain fabric, air gap, glass, and air. The local reflection coefficient under vertical polarization (consistent with the experimental configuration) is expressed as:

$$\rho_v(y) = \frac{n_1 \cos(\theta_1) - n_2 \cos(\theta_2)}{n_1 \cos(\theta_1) + n_2 \cos(\theta_2)} \tag{15}$$

where $n_1$ and $n_2$ denote the refractive indices of air and the curtain layer, respectively. The parameter $\theta_1$ represents the local incident angle $\theta_{loc}(y)$ in air, and $\theta_2$ represents the corresponding transmission angle in the curtain layer, determined by Snell's law $n_1\sin\theta_1 = n_2\sin\theta_2$. Although (15) is written in simplified Fresnel form for clarity, $\rho_v(y)$ is evaluated numerically via a coherent multilayer transfer-matrix formulation that accounts for multiple internal reflections and phase accumulations within the layered structure [50].

The angular scattering response is obtained using a physical-optics formulation, wherein the scattered field toward observation angle $\theta_s$ is calculated by coherently integrating contributions from all illuminated surface elements over the effective aperture:

$$R_v(\theta_s) = \frac{\left|\int_{-L/2}^{L/2} \rho_v(y) A(y) \exp(i\Delta\Phi(y)) dy\right|^2}{\left|E_{ref}\right|^2} \tag{16}$$

where $\Delta\Phi(y)$ accounts for the geometry-induced optical path difference and $A(y)$ accounts for the Jacobian between the parameter coordinate $y$ and the surface arc length $ds$. In 1D, $ds=\sqrt{\sec(\alpha(y))}\,dy$, so we set $A(y)=\sqrt{\sec(\alpha(y))}$ when integrating field contributions over the surface. The reference field $E_{ref}$ corresponds to the specular reflected field from a planar reference configuration ($A_c$ = 0) with identical multilayer composition and illuminated aperture, such that normalization emphasizes the curtain-induced angular redistribution.

Fig. 7(e) presents the calculated angular scattering distributions at 140 GHz and 220 GHz for a curtain-air-window structure using a cosine ripple as a first-order geometric proxy for folds, with ripple amplitude $A_c$ = 5 cm and spatial period $G_c$ = 25 cm. These dimensions were estimated independently from photographs and physical measurement of the naturally hanging curtains in the conference room, and were not adjusted to fit the measured angular response; the $\sigma$ comparison in Table V is therefore a prediction at fixed geometric inputs rather than a fit. For other frequencies, the effective dielectric parameters and fold dimensions relative to the wavelength should be re-evaluated before applying the PO model, because its validity depends on whether the folds can still be locally approximated by tilted surface elements. The model predicts strong angular redistribution over a wide range because the centimeter-scale folds introduce large local surface-normal variations while their geometric dimensions remain much larger than the wavelength. In practice, real curtains exhibit fold-to-fold variability and multi-scale perturbations, which further smear out any discrete periodic features; this explains why the measured distributions do not show stable grating orders [49]. This model prediction is validated by the experimental observations in Fig. 7(b) and Fig. 7(d), where similarly broad angular spreading without pronounced periodic features is consistently observed for the "Curtain-close" configuration.

TABLE V
QUANTITATIVE COMPARISON OF ANGULAR-SPREADING RESPONSES IN THE "CURTAIN-CLOSE" CONFIGURATION

| Frequency | $\sigma_{measure}$ | $\sigma_{model}$ | Signed error | Relative error | $NMSE_{dB}$ |
|---|---|---|---|---|---|
| 140 GHz | 29.3° | 26.7° | -2.6° | 8.8% | 0.29 |
| 220 GHz | 31.9° | 24.6° | -7.3° | 22.8% | 0.42 |

To further quantify the consistency between the measured and calculated angular-spreading behavior, the calculated data in Fig. 7(e) at $\theta_i$ = 45° for 140 and 220 GHz were extracted and compared with the measured data obtained from the corresponding angular regions in Fig. 7(a)-(d), which were defined as the positive excess components of the "Curtain-close" response relative to the "Curtain-open" reference in the linear power domain. A local angle-of-departure (AoD) integration window of $\theta \in$ [44.5°, 45.5°], centered at 45°, was used to reduce single-point sampling uncertainty. The measured and calculated profiles were normalized over the same angular range of 0°-90°. Since the folded-curtain response is characterized by distributed angular spreading rather than stable, uniquely traceable peak positions, the RMS angular spread $\sigma_x$ was adopted as a distribution-based quantitative metric [51]:

$$\sigma_x = \sqrt{\frac{\sum_{\varphi\in\Omega_\varphi} (\varphi - \bar{\varphi}_x)^2 p_x(\varphi)}{\sum_{\varphi\in\Omega_\varphi} p_x(\varphi)}} \tag{17}$$

where $x \in \{measure, model\}$ denotes the measured or calculated response, $p_x(\varphi)$ is the corresponding angular power distribution. Here, $\varphi$ represents the angle-of-arrival (AoA) for the measured profile and the corresponding observation angle for the calculated profile, and $\Omega_\varphi$ denotes the common angular range of 0°-90°. $\bar{\varphi}_x$ is the power-weighted mean angle, $\sigma_x$ is the RMS angular spread. $\varepsilon_\sigma$ denotes the relative error in this metric between the calculated and measured responses. The signed error is defined as $\sigma_{model}$ - $\sigma_{measure}$, where a negative value indicates an underestimation of the measured angular spread. As summarized in Table V, at 140 and 220 GHz, $\sigma_{measure}$ is 29.3° and 31.9°, respectively, while $\sigma_{model}$ is 26.7° and 24.6°, yielding signed errors of -2.6° and -7.3° and relative errors of 8.8% and 22.8%. In addition, the normalized mean-square error $NMSE_{dB}$ was calculated from the same extracted angular responses after converting the normalized angular-power profiles to a peak-referenced dB scale, yielding values of 0.29 and 0.42 at 140 and 220 GHz, respectively.

These results indicate that the model reproduces the overall angular-spreading scale and trend induced by the folded curtain, while systematically underestimating the absolute spread (signed errors of -2.6° and -7.3° at 140 and 220 GHz). This underestimation, and its growth with frequency, follows from the idealized surface representation rather than from a breakdown of the scattering mechanism. The PO calculation represents the curtain by a single-period cosine profile ($A_c$ = 5 cm, $G_c$ = 25 cm), whereas real hanging curtains present fold-to-fold variations in period, amplitude, and local slope, together with finer-scale surface perturbations. These unmodeled multi-scale features radiate over a wider angular range than a single deterministic period can reproduce, which accounts for the sign of the deviation. The larger deviation at 220 GHz arises from two compounding factors. First, at the shorter wavelength ($\lambda \approx$ 1.36 mm versus $\approx$ 2.14 mm at 140 GHz), a given sub-centimeter fold irregularity subtends roughly 1.6 times the phase variation, so the unmodeled fine-scale structure contributes proportionally more scattered energy that the fixed centimeter-scale idealization omits. Second, the higher antenna directivity at 220 GHz reduces spatial averaging over fold periods within the illuminated footprint, increasing the sensitivity of the measured response to the specific illuminated fold realization and to residual room-dependent multipath. Furthermore, because part of the curtain-induced scattered energy falls outside the finite angular-coverage window, the measured $\sigma$ is itself a conservative lower-bound estimate of the true angular spread; the deviation between the model and the actual channel is therefore no smaller than the reported model - measurement difference, and the PO result should be read as a lower-bound estimate of the angular-spreading scale and trend rather than a deterministic room-scale prediction. Future refinement could incorporate room-scale ray tracing to better account for scenario-dependent propagation and measurement-geometry effects. Thus the present PO calculations mainly serve to interpret the broad angular redistribution caused by folded curtains and to compare the overall angular-spreading trend.

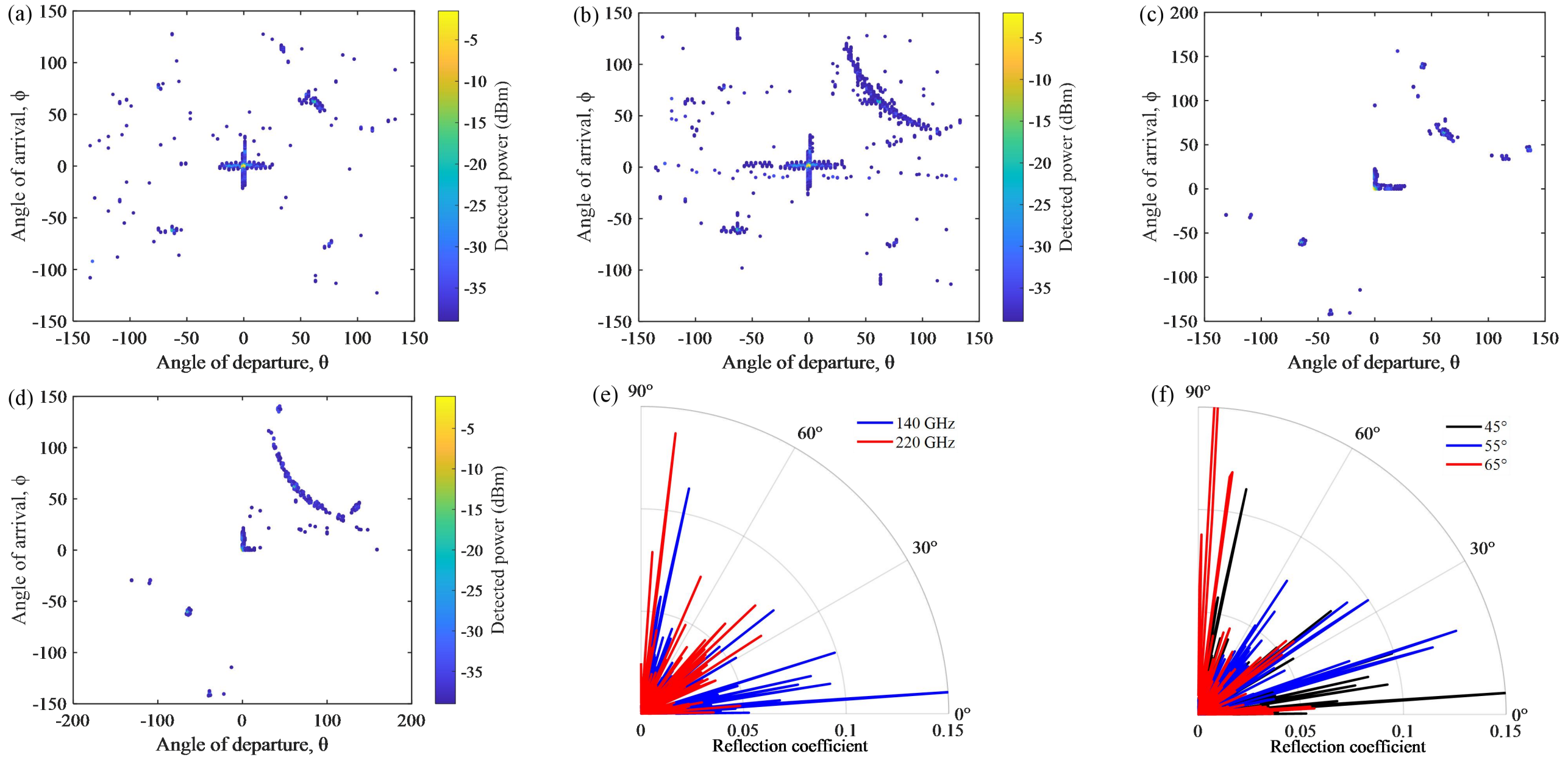

Fig. 7. (a) Measured scattering distribution at 140 GHz for the “Curtain-open” configuration. (b) Measured scattering distribution at 140 GHz for the “Curtain-close” configuration. (c) Measured scattering distribution at 220 GHz for the “Curtain-open” configuration. (d) Measured scattering distribution at 220 GHz for the “Curtain-close” configuration. (e) Calculated reflection coefficient at 140 GHz and 220 GHz for the “Curtain-close” configuration. (f) Calculated reflection coefficient as a function of incidence angle for the “Curtain-close” configuration at 140 GHz.

To further substantiate the structure-induced scattering mechanism, controlled experiments with specific curtain-fold configurations were conducted. As detailed in Fig. S4 of the Supplementary Material, scattering measurements at 113, 140, and 170 GHz were compared against those obtained for a flat curtain under identical angular configurations. At each frequency, transitioning from flat to folded curtain configuration significantly suppresses the specular reflection component at 45° (corresponding to the incidence angle), while generating additional scattering detectable over a broad angular range without stable periodic features. These results corroborate the model predictions. Meanwhile, detected power at 135° (corresponding to the LoS transmission direction) remains consistently high, indicating that the folded structure primarily redistributes reflected power over a wide angular range rather than attenuating overall transmission. The controlled folds (period $\approx$ 2 cm, amplitude $\approx$ 1.5 cm; Fig. S4) and the room-scale folds represented in the PO model (period $\approx$ 25 cm, amplitude $\approx$ 5 cm) differ by roughly an order of magnitude in fold scale, yet both produce broad angular spreading without stable periodic features. The observed redistribution is therefore a folding-induced structural effect that persists across the measured fold-scale range, rather than a property of a single fold-parameter set. A per-parameter uncertainty quantification analogous to the pine case is not available for the curtain, because the controlled and room-scale datasets sample two fold configurations rather than a continuously varied fold geometry. The robustness is accordingly stated as bounded over this measured range. The dominant uncertainty for the curtain response is the fold geometry and the Tx-curtain-Rx configuration, consistent with the larger model - measurement deviation at the higher frequency discussed above.

Since incidence angle varies continuously during full-angle indoor scans, additional calculations examining different incidence angles are presented in Fig. 7(f). Although detailed angular distributions depend on incidence angle, a common characteristic emerges: non-negligible scattered energy remains observable across a wide angular range for all configurations examined. These findings suggest that material- and structure-induced scattering mechanisms can enrich angular scattering and may provide additional propagation opportunities for indoor THz communication systems.

## V. Conclusion

As wireless communication systems progressively extend into THz bands, understanding the intricate interactions between THz channels and realistic environments becomes increasingly critical for system design and deployment. This work has systematically investigated structure-induced surface scattering phenomena at THz frequencies through coordinated experimental measurements and theoretical modeling. The dielectric characterization of six common indoor surfaces revealed substantial variations in electromagnetic properties, with the pine wood sample exhibiting particularly notable structural inhomogeneity arising from its periodic earlywood-latewood arrangement. A beam-propagation modeling (BPM) framework was developed to quantitatively link this internal periodic structure to the observed angular scattering characteristics, demonstrating that the dominant angular features are primarily governed by the effective period, tilt angle, and duty cycle of the growth-layer structure.

For material-covered surface configurations, the investigation revealed that thin dielectric covering layers can fundamentally alter the electromagnetic response through coherent thin-film interference effects, with the leather-covered aluminum surface exhibiting pronounced frequency-dependent reflection suppression near 170 GHz. Furthermore, the wallpaper-covered rough surface measurements demonstrated that dielectric coatings can redistribute the angular scattering spectrum, enhancing specular reflection while suppressing diffuse components from the underlying rough substrate. Wide-angle indoor measurements in a conference room showed that folded curtains broadened the measured angular distribution. At 140 GHz, approximately 87.5% of the total positive increase in detectable sample counts occurred within ±35° of the original specular direction. The physical-optics model captures the overall angular spreading trend but underestimates the measured RMS angular spread, with a larger discrepancy at 220 GHz. These findings collectively indicate that surface material properties, structural configurations, and environmental geometry jointly govern THz channel characteristics, and that intrinsic scattering from existing surfaces may provide additional propagation opportunities for NLoS THz scenarios.

## Appendix

With the model parameters specified in Section II-C, this Appendix summarizes the numerical implementation of the BPM calculation.

At the sample input plane, the incident field was represented by a normalized Gaussian amplitude envelope to account for the finite main-lobe illumination

$$E_{in}(x,y) = C_0 \exp\left[-\frac{x^2+y^2}{w_0^2}\right] \tag{18}$$

where $C_0$ is a normalization constant, and $w_0$ was determined from the measured main-lobe half-power beamwidth (HPBW).

To avoid artificial truncation of the incident field, a padded transverse computational window was used rather than a physical sample aperture. In the calculation, $L_x = L_y = 200$ mm, and $N_x = N_y = 1024$, giving $\Delta x = \Delta y \approx 0.196$ mm. The output angular spectrum is therefore sampled at $\Delta(\sin\theta_s) = \lambda/L_x \approx 0.0107$ ($\approx 0.61°$) near broadside at 140 GHz. The calculated angles in Table II and Table S-II lie on this grid. The BPM calculations and corresponding measurements were conducted at 140 GHz, consistent with Fig. 2. The free-space wavenumber $k_0$ enters the angular-spectrum operator of (5) and the air-side angle mapping $k_x = k_0 \sin\theta_s$. The sample thickness $d = 3$ mm was divided into $N_z = 64$ uniform propagation steps, giving $\Delta z \approx 0.0469$ mm. This calculation requires approximately $2N_z + 1 = 129$ two-dimensional FFT/inverse FFT operations, including the final far-field transform, with a computational complexity of $O(N_x N_y N_z \log(N_x N_y))$ and memory requirement of $O(N_x N_y)$.

The tilted periodic refractive-index modulation was then implemented according to (1) and (4), with $z_m$ denoting the midpoint of the $m$-th propagation slice where the local perturbation was evaluated.

For each propagation slice, the field was first updated by the local material operator defined in (6), as

$$E_m^+(x,y) = E_m^-(x,y) M(x,y,z_m) \tag{19}$$

where $E_m^-(x,y)$ and $E_m^+(x,y)$ denote the fields before and after the material modulation step, respectively.

Diffraction over one propagation step was applied in the angular-spectrum domain using the operator defined in (5):

$$E_{m+1}^-(x,y) = \mathcal{F}^{-1}\{\mathcal{F}[E_m^+(x,y)] H(k_x,k_y)\} \tag{20}$$

where $E_{m+1}^-(x,y)$ denotes the field at the beginning of the next propagation slice. The Fourier transforms were implemented in the transverse $x$-$y$ plane.

Finally, after the last propagation step, the output field was transformed into the angular-spectrum domain by a two-dimensional Fourier transform. To be consistent with the receiver scanning plane, the calculated angular response was extracted from the $k_y = 0$ cut, with the propagating components mapped to the air-side angle by $\theta_s = \arcsin(k_x / k_0)$.